\documentstyle[epsf,aas2pp4]{article}

\makeatletter
\def\tablecomments#1{\@temptokena={\vspace*{0.5ex}{%
\parbox{\pt@width}{\hskip1em{\sc Note.---}#1}\par}}%
\@temptokenb=\expandafter{\tblnote@list}
\xdef\tblnote@list{\the\@temptokenb\the\@temptokena}}
{\pt@width\wd\pt@box\center\leavevmode\box\pt@box\spew@ptblnotes%
\typeout{Page \the\pt@page\space of table \thetable\space has been set to
width \the\pt@width\space with \the\pt@nlines\space lines per page}%
\endcenter\end@float}
\newenvironment{deluxetable*}[1]{\def\pt@format{\string#1}%
\set@tblnotetext\global\pt@ncol=0\global\pt@column=0\global\pt@page=1%
\def\pt@addcol{\global\advance\pt@ncol by\@ne}%
\def\startdata{\pt@line=0\pt@calcnlines%
\ifdim\pt@width>\z@\def\@halignto{to \pt@width}\else\def\@halignto{}\fi%
\let\fnum@table=\fnum@ptable\set@mkcaption%
\@dblfloat{table}[t]\caption{\pt@caption}\vspace*{-4ex}%
\setbox\pt@box=\pt@tabular{\pt@format}\pt@head}}%
{\pt@width\wd\pt@box\center\leavevmode\box\pt@box\spew@ptblnotes%
\typeout{Page \the\pt@page\space of table \thetable\space has been set to
width \the\pt@width\space with \the\pt@nlines\space lines per page}%
\endcenter\end@dblfloat}

\def\thebibliography{\subsection*{REFERENCES}
\list{}{\labelwidth3em\leftmargin\labelwidth\labelsep\z@\parsep\z@
\itemsep\z@\itemindent-3em\usecounter{enumi}}
\def\refpar{\relax}
\def\newblock{\hskip .11em plus .33em minus .07em}
\sloppy\clubpenalty4000\widowpenalty4000
\sfcode`\.=1000\relax}
\makeatother
\def\new{PSR J2051$-$0827}
\def\orig{PSR B1957$+$20}
\def\tera{PSR B1744$-$24A}

\newcommand{\lta}{\raisebox{-0.6ex}{$\,\stackrel
{\raisebox{-.2ex}{$\textstyle <$}}{\sim}\,$}}

\def\etal{et al. }

\begin{document}
\title{The Orbital Evolution and Proper Motion of \new} 
\author{B. W. Stappers\footnote{Mount Stromlo and Siding Spring
Observatories, Institute of Advanced Studies, ANU, Private Bag, Weston
Creek, ACT 2611, Australia.}, 
M. Bailes\footnote{Astrophysics and Supercomputing, Swinburne University of
Technology, PO Box 218, Hawthorn, Victoria, 3122, Australia.}, 
R. N.  Manchester\footnote{Australia Telescope National Facility, CSIRO, PO
Box 76, Epping, NSW 2121, Australia.}, 
J. S. Sandhu\footnote{California Institute of Technology, MS 105-24,
Pasadena, CA 91125.}, 
M. Toscano\footnote{Physics Department, University of Melbourne, Parkville, Vic 3052, Australia.}}


\begin{abstract}
We have carried out high-precision timing observations of
the eclipsing binary \new\ in the 3.3 years since its discovery. These
data indicate that the orbital period is decreasing at a rate of
$\dot{P}_{b}$ = ($-11\pm$1)$\times$10$^{-12}$. If secular, this
orbital period derivative implies a decay time for the orbit of only
25\,Myr which is much shorter than the expected timescale for ablation of
the companion. We have also measured the proper motion of the pulsar
to be 5$\pm$3\,mas\,yr$^{-1}$.  Assuming the pulsar is at the
dispersion-measure distance this implies a very slow transverse
velocity $v_t$=(30$\pm$20)\,km\,s$^{-1}$. This combination of low
velocity and short orbital period argue against formation of the
system in the standard manner and we discuss the implications for its
evolutionary history.
\end{abstract}
\section{Introduction}

The discovery of the millisecond pulsar PSR J2051--0827 (\cite{sbl+96}) as
part of the Parkes survey of the southern sky for low-luminosity and
millisecond pulsars (MSPs) (\cite{mld+96,lml+98}) brought to five the number
of pulsars which are eclipsed by winds from low-mass ($\lta$ 0.1\,M\sun)
companions. It is only the second such system that is not a member of a
globular cluster and has the second-shortest orbital period of any
binary-radio pulsar, $P_{b}$=2.38\,h. The duration of the eclipse,
$\sim$10\% of the orbital period, and the delays at the eclipse egress and
ingress indicate that there is material, probably originating from a wind
driven from the companion, well beyond the companion's Roche lobe.  Thus,
this system joins both \orig\ (\cite{fst88}) and \tera\ (\cite{lmd+90}) as
neutron stars which are in the process of ablating their companions and
possibly becoming isolated MSPs (\cite{rst89a,bv91}).

Pulse times of arrival are extremely sensitive to changes in a pulsar's
local environment and along the propagation path. Timing measurements of
\new\ have already shown that there is structure in the electron column
density in the eclipse region (\cite{sbl+96}) and can provide valuable
information on the nature of the eclipses (\cite{sbl+98}).  The eclipsing
behaviour and circularity of the orbit suggest that relativistic effects are
not going to have an observable effect on the timing. However, measurements
of the changes in the orbital period for both \orig\ (\cite{rt91a,aft94}) and
\tera\ (\cite{nt96}) suggest that the effects of either the tidal influence
of the companion or the ablation process itself may be evident in the timing
data. If the pulsar has a reasonable transverse velocity then, given its
distance, $\sim$1.3\,kpc, the proper motion would also be apparent.

\section{Pulsar Timing Analysis}

Since its discovery, observations of \new\ have been made as part of our
regular high-precision timing program at Parkes (e.g. \cite{bbm+97}). Timing
data have been obtained approximately every three weeks, although there are
some large gaps resulting from unavailability of the
telescope. Approximately 2300 valid pulse times of arrival (TOAs), at centre
frequencies near 436, 660, 1520, 1940 and 2320\,MHz, have been obtained
using a filterbank system and/or the Caltech Fast Pulsar Timing Machine
(FPTM). The FPTM is based on an autocorrelator capable of very fast sampling
(\cite{nav94}). Two filterbank systems are used; at the two lower
frequencies, a 2$\times$256$\times$0.125\,MHz filterbank detects orthogonal
linear polarizations, while at the higher frequencies a
2$\times$64$\times$5\,MHz filterbank detects orthogonal circular
polarizations. Details of the filterbank-based observing systems can be
found in Manchester \etal (1996), and the correlator-based system is
described in Navarro (1994)\nocite{nav94}.

The filterbank data were reduced offline. A mean pulse profile for each
frequency channel was produced by folding at the topocentric period of the
pulsar. To compensate for the dispersive delay these profiles were
transformed to the Fourier domain, phase shifted, transformed back to the
time domain and summed. In the FPTM the down-converted signal is 2-bit
digitised and autocorrelation functions (ACFs) computed. Bandwidths were
typically 128\,MHz and 32\,MHz for observations above and below 1\,GHz,
respectively. The ACFs were hardware-integrated at the apparent pulsar
period for either 60 or 90\,s. After being transferred to a Sun Sparc-20
computer the data were calibrated, dedispersed and summed.

The typical integration time required to detect the pulsar is 60--90\,s,
although scintillation caused by the interstellar medium means that much
longer integration times are sometimes required to obtain an acceptable
signal-to-noise ratio. Each integration was optimized to maximize orbital
phase coverage with individual TOAs, while preserving a minimum
signal-to-noise ratio of $\sim$12. The observed profiles were then
cross-correlated with a standard pulse template at each frequency to
determine the TOA. Separate templates were used for the higher resolution
correlator data and no offsets between the two data sets were required to
the limit of the best fitting residuals. These data were then fitted using the
{\tt TEMPO} timing analysis package (\cite{tw89}) with the JPL DE200
ephemeris (\cite{sta82}). Binary parameters were fitted using the timing
model of Blandford \& Teukolsky (1976)\nocite{bt76}.

\section{Parameters}

A binary timing model which included the pulsar position, dispersion
measure, rotational period and period derivative, orbital period, projected
semi-major axis of the orbit, and orbital period derivative was used to
generate the best fitting parameters shown in Table \ref{params}. Data which
lie between orbital phases 0.2 and 0.35 were discarded from all fits because
of the influence of the excess column density in the eclipsing region on
these TOAs. This range is narrower than that rejected by Stappers \etal
(1996) as the eclipse boundaries are now better defined. In all fits, both
the eccentricity and the longitude of periastron were held fixed at zero to
determine the remaining parameters. The eccentricity limit was determined by
fitting simultaneously for the eccentricity and the remaining parameters for
various input values of the longitude of periastron.

\begin{figure*}[htb]
\epsfbox{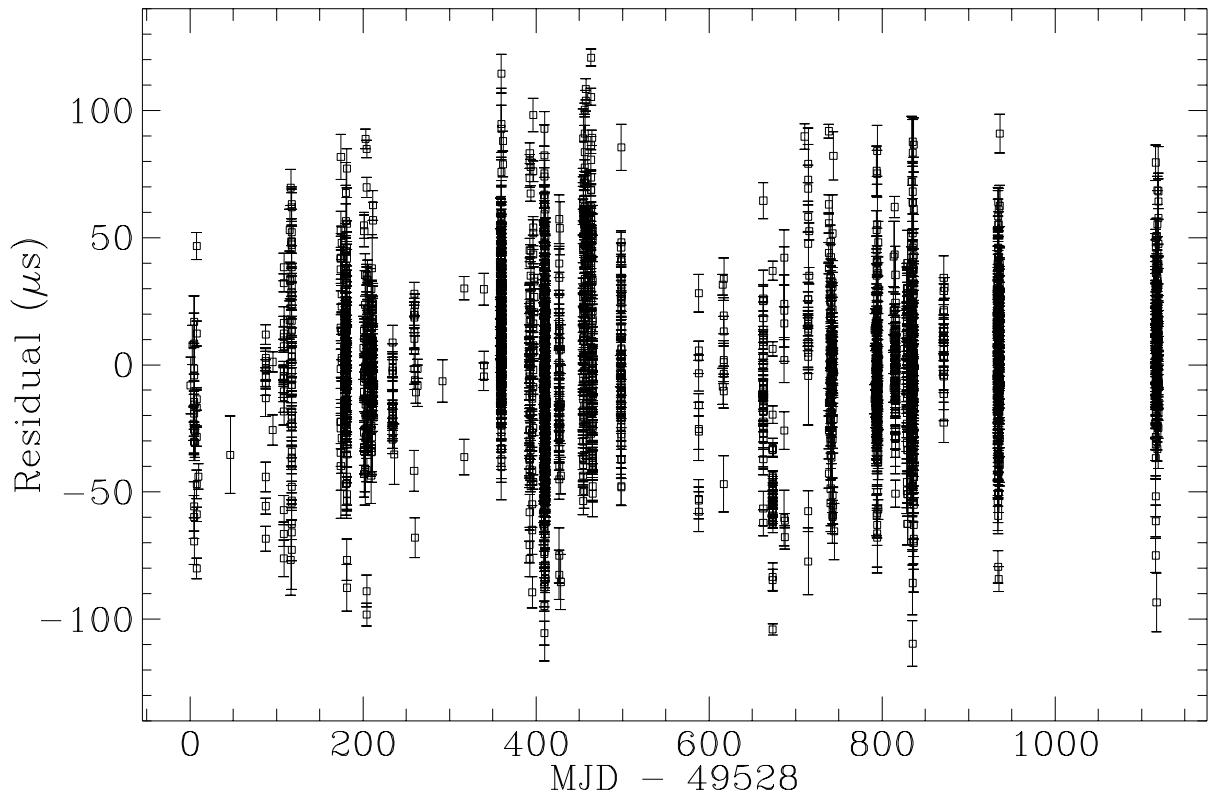}
\caption{Post-fit timing residuals for $\sim$2300 timing measurements of
  \new\ at a number of different frequencies (see text) for the parameters
  in Table 1.}
\label{resid}
\end{figure*}

The residuals for this best fit are shown in Figure \ref{resid}. There is
unmodelled timing noise on timescales ranging from less than an orbit to a
few orbits. The intra-orbit variations can be attributed to the low
signal-to-noise of some profiles. The errors are from the formal fit and
assumed that the errors were completely random. In practice narrow-band
interference introduces ripples into the pulse profile which lead to
underestimates of the true error. There are no trends observed in these
changes and a sufficiently large number of orbits have been observed for
orbital characteristics to remain unaffected. Changes on the timescale of a
few orbits may be due to changes in the electron column density. Variations
on longer time scales similar to those seen for \orig\ (\cite{rt91a,aft94})
and \tera\ (\cite{nt96}) are not present. A formal fit for a period second
derivative to the timing data of \new\ yields
$\ddot{P}$=(4$\pm$2)$\times10^{-29}$s$^{-1}$, suggesting that mass motions
in the \new\ system do not greatly affect the rotation of the pulsar.

If the transverse velocity of a pulsar is sufficient, then the contribution
of the apparent acceleration along the line of sight to the derivatives of
the orbital and rotational period of the pulsar is significant
(\cite{shk70}). This is especially the case for MSPs which have rotational
spin-down rates smaller than those of the long-period pulsars. Corruption of
the rotational period derivative means all properties of MSPs which are
derived from it, such as the characteristic age, $\tau_c = P/2\dot{P}$, and
the pulsar spin-down energy, $\dot{E}_p = 4\pi^{2}I\dot{P}/P^{3}$ (I =
10$^{45}$\,g\,cm$^{-2}$), are only provisional until the velocity, or an
upper limit, can be determined. Modelling the lightcurve of the companion to
\new\ has shown that at least 30\% of the pulsar's spin-down energy may be
required to heat the companion star (\cite{svlk98}); this fraction would
increase to nearly 50\% for a proper motion of 16\,mas\,yr$^{-1}$
($v_{t}\sim$100\,km\,s$^{-1}$).  However, as shown in Table \ref{params},
our current measurement of the composite proper motion of \new\ is
5$\pm$3\,mas\,yr$^{-1}$. Assuming the dispersion-measure distance to the
system, this implies a remarkably slow transverse velocity of only
$v_t$=(30$\pm$20)\,km\,s$^{-1}$. Thus the contribution to the period
derivative is 3.4$\times$10$^{-22}$, or just 3\% of the measured period
derivative.

A significant result from these timing observations is that the orbital
period of \new\ is decreasing at a rate $\dot{P}_{b}$ =
($-11\pm$1)$\times$10$^{-12}$. This $\dot{P}_{b}$ is some two orders of
magnitude greater than the contribution expected from general relativistic
effects, $\dot{P}_{b}$ = ($-3\pm$1)$\times10^{-14}$, and the possible
influence of the Shklovskii term is negligible. If the orbital period
derivative were constant then the orbital decay time would be only
25\,Myr. However, observations of \orig\ indicate that its orbital period
derivative varies, and even changes sign, on quite short time scales
(Arzoumanian \etal 1994). The best fit obtained for a constant orbital
period has an r.m.s $\sim$35\,$\mu$s, somewhat greater than for the fit in
Table \ref{params}.

If the orbital period is decreasing, the pulsar should arrive at the
ascending node earlier than we would predict from the constant orbital
period model. Data presented in Figure \ref{resid}, except those obtained in
1997 July, were split into three, approximately equal, time-interval
groups. After re-assigning the epoch of ascending node to the middle of each
data set, they were fitted for orbital and spin parameters. The position was
held constant for all fits as none of the data sets spanned a full year, and
the proper motion is small. The measured phase shifts are found to be
consistent with those predicted by the measured orbital period derivative.
Thus confirming the orbital period of the system is presently
decreasing. There is insufficient data to fit for higher-order variations in
the orbital period.

\section{Evolution: Past}

The orbital period of \new\ puts it just above the upper edge of the
low-mass x-ray binary (LMXB) period gap. It is one of only four binary
pulsars with a period less than 4\,h. It is interesting to note that of
these four binary pulsars, only J1910+0004 (\cite{dma+93}) shows no strong
evidence of eclipsing behaviour, and it is probably too weak to preclude the
possibility that it is also an eclipsing system.  Thus at least three out of
the four systems are eclipsed, more than would be expected, naively, if
eclipses in these systems required inclination angles close to
90$^{\circ}$. This is further evidence that there must be material which
extends well beyond the companion's Roche lobe. The eclipsing phenomenon is
perhaps inevitable for such short-period binaries.

The proper motion and dispersion-measure-derived distance of \new\ indicate
its transverse velocity is very low (30 km\,s$^{-1}$). Unless there is an
unusually large radial velocity component, the space velocity will also be
small. There are presently ten MSPs with estimates of their proper motions
and they indicate a mean space velocity of $\sim$100\,km\,s$^{-1}$. This is
significantly less than the mean velocity of the long-period pulsars
(e.g. Lyne \& Lorimer 1994\nocite{ll94}). Simulations by Cordes \& Chernoff
(1997)\nocite{cc97} and Ramachandran \& Bhattacharya (1997)\nocite{rb97}
show that such a low mean velocity is expected for the MSP population as
progenitor systems that receive high kick velocities will disrupt during the
supernova. Thus, the low velocity of \new\ is consistent with its proximity
to the disk (z=650\,pc) but is problematic when considering its past
evolution.

In standard evolutionary models, low-mass binary and isolated MSPs are
descended from LMXBs (\cite{acrs82,bv91}). LMXBs typically have orbital
periods of a few days before the supernova explosion of the primary. The
simulations of Tauris \& Bailes (1996)\nocite{tb96} show it is almost
impossible for such initially compact systems to generate binary MSPs with
orbital periods less than a day and spatial velocities less than
50\,km\,s$^{-1}$. This is because the shell of the exploding star removes a
large amount of momentum from the system when the neutron star is formed,
and the binary recoils. The closer the system at the time of the explosion,
the greater the recoil velocity. Thus we expect the shortest orbital-period
MSPs to have the greatest velocities.  While the transverse velocity,
$v_t$$\approx$220\,km\,s$^{-1}$ of \orig\ (Arzoumanian \etal 1994), is
consistent with the standard formation mechanism and the simulations of
Tauris \& Bailes (1996), the low velocity of \new\ indicates that it
probably did not form this way.

Any alternative formation process must explain the low velocity, short
orbital period, and 4.5\,ms pulse period of \new. The low velocity suggests
the orbital separation at the time of the explosion was wide. The only
reasonable option requires a massive spiral-in since the formation of the
neutron star. Mass transfer from a more massive to a less massive body leads
to spiral-in, but it must occur slowly enough for the neutron star to be
spun up to a millisecond period. Thus the companion cannot have been too
massive. Our ideal progenitor system is therefore an intermediate-mass
binary with an orbital period of a few months and a secondary in the mass
range 2-4 M$_\odot$.  There is enough mass for the system to avoid
disruption when the primary explodes, a sufficiently long orbital period to
ensure a low runaway velocity, the correct mass-ratio for eventual spiral-in
of the neutron star into the secondary core, and enough time to spin up the
pulsar.

This alternative evolutionary process may explain the low velocity of \new\
but nominally its end product is a MSP and a massive white dwarf in a
circular orbit. If the remaining core mass after the common-envelope phase
was sufficiently small, then mass transfer to the neutron star may not have
become unstable. The final spiral-in phase would have been avoided and the
system may have proceeded instead to a compact LMXB-like phase. This
scenario is attractive as the evolution of the system would then be similar
to that discussed by Ruderman \etal (1989)\nocite{rst89a} and would
naturally explain the present very low-mass companion.

Alternatively, if the neutron star was formed via accretion-induced collapse
of a massive white dwarf (e.g. Bailyn \& Grindlay 1990\nocite{bg90}) then
the velocity of the system would also be lower as less mass is lost during
formation of the neutron star. However, it would also require the explosion
to be symmetric, otherwise the system would still be expected to have a large
velocity (\cite{tb96}) and would require the resultant pulsar to ablate most of
the mass from the companion.

\section{Evolution: Future}

Initial ablation of the companion in eclipsing and isolated systems is
thought to occur during an LMXB phase when a $\gamma$-ray flux may be
generated by the interaction of the neutron star's magnetic field and the
accretion disk. This flux causes a wind to be driven from the companion star
by heating its outer layers (\cite{rste89,krst88}). Evaporation of the
companion may eventually cause the accretion to cease and the spun-up pulsar
would then be able to turn on. The pulsar may then continue to ablate the
companion through its relativistic wind (\cite{rst89a}). Following the
simple calculation of Bhattacharya \& van den Heuvel (1991), we relate the
rotational energy of the pulsar to the binding energy of the companion to
determine if the pulsar can ablate its companion. Assuming the ablation
process is 1\% efficient, i.e., 1\% of the incident pulsar spin-down energy
is available to drive the wind, we find \new\ will require a further
10$^{9}$\,yr to evaporate its companion. This time scale is longer than the
25\,Myr time scale for orbital decay suggesting that tidal effects may be
more important in completely destroying the companion. If MSPs live for
10$^9$ yr or more this time scale for destruction of the companion via
orbital decay is inconsistent with the observed number of eclipsing MSPs and
isolated MSPs.

The value for the orbital period derivative is much larger than that
predicted by gravitational radiation losses. Mechanisms whereby angular
momentum is lost from the system through anisotropic mass loss from the
companion's surface (\cite{bs92,eic92,bt93}) were proposed to explain the
orbital period derivative of \orig. However, they require a large mass-loss
rate which is hard to reconcile with the low electron densities measured in
the eclipse region of either \orig\ (e.g. \cite{fbb+90}) or \new\ 
(\cite{sbl+96}). Moreover they are unable to reproduce the rapid change in
sign and magnitude of the orbital period derivative that has been measured
for \orig.

These variations in the period derivative have been likened to the
quasi-cyclic variations seen in other close binaries where magnetic fields
of the companion are important (Arzoumanian \etal 1994). Based on this idea,
Appelgate \& Shaham (1994)\nocite{as94}, developed a model where the
companion's magnetic activity and wind generate a torque which prevents it
from co-rotating and results in dissipation of tidal energy. Although we
have measured only a secular change in the orbital period derivative of
\new\ at this stage, given its companion mass is similar to that for \orig,
and the orbital separation is less, we might expect it too will have a
variable orbital period derivative. Clearly if such variation in the orbital
period derivative were measured for \new, it would indicate it will live
longer than 25\,Myr and thus alleviate the birthrate problem.

The low electron column densities measured in the eclipse region for \new\
suggest that the current evolutionary phase is a long-lived one. However,
the large orbital period derivative, if it is secular, provides a mechanism
to reduce this system to an isolated MSP. The low velocity of
the pulsar is a powerful constraint on the progenitor system and favours its
having evolved from an initially intermediate mass binary.

We thank S. Johnston and the referee for helpful comments on the
manuscript. We are grateful to S. Anderson, M. Britton, A. Hughes, V. Kaspi,
A. Lyne and the Parkes Observatory staff for assistance with
observations. B. W. S. received support from an ANU PhD scholarship and the
ATNF student program.

\begin{table} 
\begin{center} 
\caption{Parameters of \new}
\begin{tabular}{ll} 
\hline \hline 
Right Ascension (J2000) &  $20^{\rm h}51^{\rm m}07\fs5130(2)$ \\ 
Declination (J2000)     &  $-08^{\circ}27\arcmin 37\farcs782(6)$ \\ 
Proper motion in R.A.   &  1(2) mas yr$^{-1}$\\
Proper motion in Decl.  &  $-$5(3) mas yr$^{-1}$\\
Epoch of Period (MJD)   &       49530.0 \\ 
Period (s)      &       0.0045086417433540(7) \\ 
Period Derivative ($\times 10^{-20}$)   &       1.272(2) \\ 
Dispersion Measure (cm$^{-3}$ pc)       &       20.7458(2) \\ 
Orbital Period (days)   &       0.0991102650(2) \\ 
Orbital Period Derivative ($\times10^{-12}$)  & $-$10.8(10) \\
Projected Semi-major Axis (lt-s)        &       0.045076(1) \\ 
Eccentricity & $<8\times10^{-5}$ \\
Epoch of Ascending Node (MJD) &  50015.422291 \\ 
R.M.S. timing residual ($\mu$s) &      32 \\
 & \\
Distance (kpc)  & 1.3 \\
Galactic Longitude (degrees) & 39.19 \\ 
Galactic Latitude (degrees) & $-$30.41 \\
Mass function ($M_{\odot}$) &  1.00108(7)$\times10^{-5}$ \\
Minimum Companion mass ($M_{\odot}$) & 0.027 \\  
\hline 
\end{tabular} 
\label{params}
\end{center} 
\end{table} 

\bibliographystyle{apj1}

\end{document}